\documentclass{elsart}
\usepackage{epsfig}
\usepackage{graphicx}
\usepackage{amssymb}
\journal{Physics Letters B}
\def\bbox#1{\mbox{\boldmath $#1$}}
\begin{document}
\begin{frontmatter}
\title{Di-neutron  elastic transfer in the $^4$He($^6$He,$^6$He)$^4$He
 reaction\thanksref{Ackn}}
\thanks[Ackn]{Research supported, in part, by the
Alexander-von-Humboldt Stiftung of Germany and Natural Science Council of
Vietnam.}

\author[INST]{Dao T. Khoa\corauthref{cor1}} and
\ead{khoa@vaec.gov.vn}
\corauth[cor1]{Corresponding author.}
\author[HMI]{W. von Oertzen\thanksref{cor2}}
\ead{oertzen@hmi.de}
\thanks[cor2]{Also Fachbereich Physik, Freie Universit\"at Berlin.}
\address[INST]{Institute for Nuclear Science and
 Technique, VAEC, P.O. Box 5T-160,\\ Nghia Do, Hanoi, Vietnam.}
\address[HMI]{Hahn-Meitner-Institut GmbH, Glienicker Str. 100, D-14109 Berlin,
 Germany.}

\begin{abstract}
Elastic $^{6}$He+$^4$He data measured at $E_{\rm c.m.}=11.6,$ 15.9, and 60.3 MeV
have been analyzed within the coupled reaction channels (CRC) formalism, with
the elastic-scattering and two-neutron ($2n$) transfer amplitudes coherently
included. Contributions from the direct (one-step) and sequential (two-step)
$2n$-transfers were treated explicitly based on a realistic assumption for the
$2n$-transfer form factor. The oscillatory pattern observed in
$^4$He($^6$He,$^6$He)$^4$He angular distribution at low energies was found to be
due to an interference between the elastic scattering and $2n$-transfer
amplitudes. Our CRC analysis shows consistently that the direct $2n$-transfer
strongly dominates over the sequential transfer and thus confirms the dominance
of 2$n-^4$He configuration over the $n-^5$He one in the $^6$He wave function.
This result suggests a strong clusterization of the two valence neutrons and
allows, therefore, a reliable estimate for the \emph{di-neutron} spectroscopic
amplitude.
\end{abstract}

\end{frontmatter}

With $^{6}$He beams becoming available with hight intensity and good resolution,
this Borromean nucleus is now one of the most studied unstable nuclei. It is
well established that $^{6}$He consists of an inert $^4$He core and two valence
neutrons, with the two-neutron ($2n$) separation energy of 0.975 MeV
\cite{Zhu93}. The Borromean binding mechanism implies that the two valence
neutrons cannot bind to the $^4$He core separately but only as a pair and a
strong di-neutron correlation is necessary for the formation of the $2n$-halo
\cite{Nie01,Oga99}. However, a direct experimental confirmation about the
neutron-neutron correlation in this nucleus is not a simple task. For example,
from the measurement of the total reaction (or interaction) cross section (see
Ref.~\cite{Oz01} and references therein) one can only deduce the long tail of
the $^{6}$He ground-state (g.s.) density which is due to the low binding energy.
Elastic and inelastic scattering of $^6$He beams on proton target
\cite{Lap01,Lag01} is more informative for testing the halo tail, especially
when measured with high precision over a large angular range \cite{egelhof}, but
also cannot provide any information on the di-neutron correlation in $^6$He.

In difference to the above experiments, neutron transfer reaction induced by
$^6$He beams can provide us with valuable information about the two valence
neutrons in $^6$He \cite{Oga99,Tim00}. In the same way as a single-nucleon
transfer reaction delivers reliable estimate for the spectroscopic factor of a
single-nucleon configuration, it is expected that the $2n$-transfer reaction
induced by $^6$He beam will also provide the spectroscopic information about
$2n$ wave function. For this purpose, elastic $^4$He($^6$He,$^6$He)$^4$He
reaction has been measured at $E_{\rm c.m.}=60.3$ MeV by Ter-Akopian {\sl et
al.} \cite{Ter98} and the data show a rise of the elastic cross section at
backward angles which is likely due to the elastic $2n$-transfer process
\cite{voebohlen}. An analysis of these data in the distorted wave Born
approximation (DWBA) by Oganessian {\sl et al.} \cite{Oga99} has shown that from
the two configurations (``di-neutron" and ``cigar") predicted for $^6$He
\cite{Zhu93} the di-neutron configuration is dominant. It should be noted,
however, that the conclusion made in Ref.~\cite{Oga99} is meaningful only if the
direct one-step transfer dominates over the sequential transfer. The experience
with the $2n$-transfer reactions measured with heavy ions is such that the
direct and sequential transfer amplitudes are of similar magnitude, especially
at low energies (see, in particular, Sect. 16.6.5 of Ref.~\cite{Sat83}).
Therefore, the contribution of sequential two-step transfer
($^6$He,$^5$He;$^5$He,$^4$He) to the total ($^6$He,$^4$He) transfer amplitude
should be carefully investigated before making conclusion about the di-neutron
configuration in $^6$He. In addition, the 2$n$-transfer might also include an
indirect route \cite{voevitturi,Kro01} through the 2$^+$ excitation of $^6$He.

Elastic $^4$He($^6$He,$^6$He)$^4$He reaction has been measured recently at lower
energies of $E_{\rm c.m.}=11.6$ and 15.9 MeV by Raabe {\sl et al.}
\cite{Raa99,Raa03}. The most accurate are the data at $E_{\rm c.m.}=11.6$ MeV
which were measured in two separate experiments using the static $^4$He gas
\cite{Raa99} and $^4$He-implanted Al foil \cite{Raa03} as targets. Elastic
$2n$-transfer contribution in the $^4$He($^6$He,$^6$He)$^4$He reaction at
$E_{\rm c.m.}=11.6$ MeV was estimated \cite{Raa03} within DWBA, assuming the
direct transfer of a structureless $2n$-cluster in the $S$-state. It was also
shown in Ref.~\cite{Raa03} that these data are well reproduced in a simple
optical model (OM) calculation using the double-folding potential obtained by
Baye {\sl et al.} \cite{Bay96} added by a parity-dependent term to simulate
elastic $2n$-transfer, in the same way as discussed in Ref.~\cite{voebohlen}.
Although these DWBA and OM results deliver some estimate of the $2n$-transfer
strength in the $^4$He($^6$He,$^6$He)$^4$He reaction, they can be of little help
in extracting finer details about the $^6$He structure. The situation becomes
further confused by a coupled-discretized-continuum channels (CDCC) analysis of
the $^4$He($^6$He,$^6$He)$^4$He data at $E_{\rm c.m.}=11.6, 15.9,$ and 60.3 MeV
by Rusek and Kemper \cite{Rus00}, where the rise of elastic $^6$He+$^4$He
scattering cross section at backward angles is reproduced by using a weakly
absorptive optical potential (OP) and taking into account only the coupling
between the elastic scattering and breakup channels.

The purpose of our investigation is, therefore, to include a realistic reaction
mechanism into the analysis of the $^4$He($^6$He,$^6$He)$^4$He reaction and to
estimate as accurate as possible the $2n$-transfer contribution. We analyze the
elastic $^{6}$He+$^4$He data at $E_{\rm c.m.}=11.6, 15.9,$ and 60.3 MeV
consistently within the coupled reaction channels (CRC) formalism and treat
contributions from the direct (one-step) and sequential (two-step)
$2n$-transfers explicitly based on a realistic choice of the transfer form
factors. All the CRC calculations were made using version FRXY.1d of the code
FRESCO written by Thompson \cite{Tho88}. A recent CRC analysis \cite{Kro01} of
the $^4$He($^6$He,$^6$He)$^4$He reaction has been done for $E_{\rm c.m.}=60.3$
MeV only, and a consistent CRC analysis of extensive data at the three available
energies is strongly needed.

The basic ingredients of the present CRC calculation are the (diagonal) OP and
(off-diagonal) \emph{nonlocal} coupling potentials due to the one- and
two-neutron transfers. Dropping the indices of angular momenta for simplicity,
the coupling potential of the direct (one-step) $2n$-transfer between $^6$He and
$^4$He can be written as
\begin{equation}
<\Psi_{^6{\rm He}}(\bbox{r}'_{\alpha 2},\bbox{r}'_1,\bbox{r}'_2)
 \Psi_{^4{\rm He}}(\bbox{r}'_{\alpha 1})|V_{\rm direct}|
 \Psi_{^6{\rm He}}(\bbox{r}_{\alpha 1},\bbox{r}_1,\bbox{r}_2)
 \Psi_{^4{\rm He}}(\bbox{r}_{\alpha 2})>,
 \label{e1}
\end{equation}
where $\bbox{r}_{\alpha 1,2}$ and $\bbox{r}'_{\alpha 1,2}$ are the
centers-of-mass coordinates of the two $^4$He cores before and after collision,
respectively, and similarly for the coordinates $\bbox{r}_{1,2}$ of the two
valence neutrons in $^6$He. The transfer interaction is further determined, in
the post form \cite{Sat83}, as
\begin{equation}
V_{\rm direct}=V_{1-\alpha}+V_{2-\alpha}+(U_{^4{\rm He}-^4{\rm He}}-
 U_{^6{\rm He}-^4{\rm He}}),
 \label{e2}
\end{equation}
where $V_{1,2-\alpha}$ are the potentials binding each of the two valence
neutrons to the $^4$He core in $^6$He, $U_{^4{\rm He}-^4{\rm He}}$ and
$U_{^6{\rm He}-^4{\rm He}}$ are the core-core and $^{6}$He+$^4$He optical
potentials at the considered energy. For the sequential (two-step) transfer
$^{6}$He+$^4$He$\to ^{5}$He+$^5$He$\to ^{4}$He+$^6$He, one needs to determine
the (post-form) transfer interactions
\begin{equation}
 V^{(1)}_{\rm seq.}=V_{n-^5{\rm He}}+(U_{^5{\rm He}-^4{\rm He}}-
 U_{^5{\rm He}-^5{\rm He}})
 \label{e4a}
\end{equation}
\begin{equation}
 V^{(2)}_{\rm seq.}=V_{n-\alpha}+(U_{^5{\rm He}-^4{\rm He}}-
 U_{^6{\rm He}-^4{\rm He}})
 \label{e4b}
\end{equation}
for the first- and second-step transfers, respectively. We treat the $^4$He core
in our CRC calculation as a structureless particle with spin $J_{^4{\rm
He}}=0^+$, and both the direct and sequential $2n$-transfer amplitudes are
properly \emph{symmetrized} with respect to the $^4$He exchange. In such an
approximation, the transfer operators (\ref{e2}), (\ref{e4a}) and (\ref{e4b}) do
not act on the internal coordinates of $^4$He, and the coupling matrix element
(\ref{e1}) turns out to be directly proportional to the overlap $<\Psi_{^4{\rm
He}}|\Psi_{^6{\rm He}}>$ and vertex $<\Psi_{^4{\rm
He}}|V_{1\alpha}+V_{2\alpha}|\Psi_{^6{\rm He}}>$ for the direct transfer.
Similarly, the coupling terms for the sequential transfer are determined by the
overlaps $<\Psi_{^4{\rm He}}|\Psi_{^5{\rm He}}>$ and $<\Psi_{^5{\rm
He}}|\Psi_{^6{\rm He}}>$, and vertices $<\Psi_{^4{\rm He}}|V_{n-^5{\rm
He}}|\Psi_{^5{\rm He}}>$ and $<\Psi_{^4{\rm He}}|V_{n\alpha}|\Psi_{^5{\rm
He}}>$, respectively.

We discuss now our choice of the wave functions $|\Psi_{^{5,6}{\rm He}}>$. Since
g.s. of $^5$He is  a $p\frac{3}{2}$ resonance of 0.6 MeV width and unbound by
0.89 MeV, we have adopted the same \emph{quasi-bound} approximation as that used
in Ref.~\cite{Kro01} for the $p\frac{3}{2}$ valence neutron in $^5$He. This is a
reasonable approximation which produces a fast decaying tail of this $l=1$
state. The standard ``core + valence neutron" option of the code FRESCO
\cite{Tho88} was used for the coupling potentials of the sequential transfer,
where the $p\frac{3}{2}$ valence neutron in $^5$He is bound by the potential
$V_{n-\alpha}$ consisting of a central Woods-Saxon (WS) potential with
$r_0=1.35$ fm, $a=0.65$ fm and a spin-orbit term of the Thomas form
\cite{Sat83}. The WS depth was fixed to reproduce the quasi-binding energy of
0.01 MeV and strength of the spin-orbit potential was taken to be 17 MeV. Note
that such a quasi-bound approximation is used in the construction of one-neutron
transfer form factor only, and the CRC calculation includes the correct
(\emph{negative}) $Q$-value for the sequential transfer channel.

The same WS geometry, as that of $V_{n-\alpha}$ potential, was used for
$V_{n-^5{\rm He}}$ binding potential to generate $p\frac{3}{2}$ wave function
for each of the two valence neutrons in $^6$He, but with the WS depth adjusted
to reproduce the experimental one-neutron separation energy of 1.86 MeV.
Although the Borromean binding mechanism does not necessarily leads to the
single-neutron wave function with the asymptotic defined by the one-neutron
separation energy of $^6$He, this approximation has been proven to be reasonable
in the CRC analysis of proton-induced reactions on $^6$He \cite{Tim00} or in the
calculation of $^6$He g.s. density \cite{Kho04}. Then, $|\Psi_{^6{\rm He}}>$ is
modelled by a ``core + 2$n$" bound-state wave function
\begin{equation}
 |\Psi_{^6{\rm He}}>=|\Psi_{^4{\rm He}}\otimes (p\frac{3}{2})^2>
 \equiv \sum  |NL(nlJ);0^+>, \label{wf0}
\end{equation}
where $J$ is the internal angular momentum of the $2n$-cluster and $L$ is its
orbital angular momentum with respect to the $^4$He core. The relative motion of
the two $p\frac{3}{2}$ neutrons coupled to $J=0$ and 1 in the cluster frame is
described by $|nlJ>$ [see Eq.~(3.23) in Ref.~\cite{Tho88} for the explicit
expression of (\ref{wf0})]. As a result, we have taken into account all
configurations with $l=L=0$ and 1, where $l$ is the relative orbital angular
momentum between the two neutrons. These configurations were shown to give the
most important contributions to the $^6$He binding energy \cite{Fun94}. Thus,
the g.s. wave function (\ref{wf0}) consists only of two parts: $S$-wave (with
$J=L=0$) and $P$-wave (with $J=L=1$). Since (\ref{wf0}) is not a solution of a
structure model, we need to assign the amplitudes of the $S$- and $P$-waves as
accurate as possible for the CRC analysis. Given the results of the microscopic
structure study of $^6$He \cite{Zhu93,Nie01} using the hyperspherical basis
which give $P$-wave probability $P_p\approx 10-14\%$, as well as those of the
three-cluster model calculation \cite{Bay96} which give $P_p\approx 17\%$, we
have included explicitly into the CRC calculation such $S$- and $P$-wave
amplitudes that give $P_s=85\%$ and $P_p=15\%$. Note that $P$-wave does not
contribute to the ``di-neutron'' configuration because of the centrifugal
barrier and corresponds more likely to the ``cigar'' configuration \cite{Tim01}.
A probe of the $S$- and $P$-wave contributions to the $2n$-transfer cross
section is, therefore, necessary before a conclusion about the di-neutron and
cigar configurations is made.

We discuss now our choice of He-He potentials used in the coupling terms
(\ref{e2}), (\ref{e4a}), and (\ref{e4b}) for $2n$-transfer. In general,
realistic complex OP's for $^6$He+$^4$He, $^5$He+$^5$He, $^5$He+$^4$He, and
$^4$He+$^4$He systems should be used in the CRC calculation. Since the
double-folding model (DFM) \cite{Sat79} for the real part of nucleus-nucleus OP
has been proven to be quite accurate at low and medium energies, we use the
latest version of DFM \cite{Kho00,Kho01} to calculate the real OP ($V_{\rm
Fold}$) based on the CDM3Y6 density dependent interaction \cite{Kho97} and
realistic choice of the g.s. densities for $^{4,5,6}$He. Such a folding approach
was used recently by Avrigeanu {\sl et al.} \cite{avrigeanu} to successfully
predict real OP's for $^6$He+$^4$He and $^6$He+$p$ systems. The imaginary OP is
due to the coupling of the elastic channel to all nonelastic channels, and to
calculate it microscopically will be a task far beyond the scope of DFM
\cite{Sat79}, especially for a weakly bound projectile like $^6$He. Therefore,
the standard WS shape is used for the imaginary OP and the total OP is
determined as

\begin{equation}
U(R)=N_{\rm V}V_{\rm Fold}(R)+V_{\rm C}(R)+iW(R), \label{e5a}
\end{equation}
\begin{equation}
{\rm where}\ W(R)=-W\left[1+\exp
 \left(\frac{R-R_{\rm W}}{a_{\rm W}}\right)\right]^{-1}. \label{e5b}
\end{equation}

Here $V_{\rm C}$ is the Coulomb potential between a point charge and a uniform
charge distribution of radius $R_{\rm C}=1.25(A_1^{1/3}+A_2^{1/3})$ fm,
 $R_{\rm W}=r_{\rm W}(A_1^{1/3}+A_2^{1/3})$. Parameters of the WS imaginary OP
(Table~\ref{t1}) and renormalization factor of the \emph{energy-dependent} real
folded potential ($N_{\rm V}=1.15$) were chosen  to reproduce the data points at
the most forward angles, where elastic scattering dominates.

\begin{table}\small
\caption{\small WS parameters of $^6$He+$^4$He imaginary OP (\ref{e5b}) used in
the CRC analysis of $^4$He($^6$He,$^6$He)$^4$He reaction with or without the
coupling to the 2$^+$ excitation of $^6$He. $\sigma_{\rm R}$ and $\sigma_{2^+}$
are the total reaction cross section and integrated 2$^+$ inelastic cross
section, SA$_{\rm g.s.}$ and SA$_{2^+}$ are the di-neutron spectroscopic
amplitudes of $^6$He$_{\rm g.s.}$ and $^6$He$^*_{2^+}$, respectively.}
\label{t1}
\begin{tabular}{c c c c c c c c c c} \hline\small
 $E_{\rm c.m.}$ & $J_{\rm V}$ & Coupling & $W$ & $r_{\rm W}$
 & $a_{\rm W}$ & $\sigma_{\rm R}$ & $\sigma_{2^+}$ & SA$_{\rm g.s.}$ &
 SA$_{2^+}$ \\
MeV & MeV~fm$^3$ & g.s. $\leftrightarrow 2^+$ & MeV & fm & fm & mb & mb & &
\\ \hline
11.6  & 443 & No & 29.0 & 1.30 & 0.25 & 859 & 0 & 1.15 & 0 \\
11.6  &  443 & Yes & 23.0 & 1.30 & 0.28 & 801 & 53.8 & 0.95 & 1.3 \\
15.9 &  436 & No & 29.0 & 1.15 & 0.20 & 756 & 0 & 1.15 & 0 \\
60.3 &  376 & No & 9.00 & 1.25 & 0.75 & 829 & 0 & 1.00 & 0 \\
60.3 &  376 & Yes & 7.50 & 1.25 & 0.75 & 797 & 36.1 & 0.85 & 1.3 \\
\hline
\end{tabular}
\end{table}

$^4$He density was taken in the Gaussian form adopted in Ref.~\cite{Sat79},
which has been proven by a folding analysis of $\alpha$-nucleus elastic
scattering \cite{Kho01} as the most realistic. The independent-particle model
(IPM) \cite{Sat79,Sat80} (which generates each single-nucleon orbital using an
appropriate WS potential added by a spin-orbit term of the Thomas form) was used
to calculate g.s. densities of $^{5,6}$He. Since $^{6}$He can be produced by
picking up a proton from $^{7}$Li \cite{Oga99}, we have used the same
$s\frac{1}{2}$ binding potential as that used for $^{7}$Li by Satchler
\cite{Sat79} ($r_0=1.25$ fm, $a=0.65$ fm for $s\frac{1}{2}$ neutrons and protons
which are bound by $S_n=25$ MeV and $S_p=23$ MeV, respectively), but with the
recoil effect \cite{Sat80} properly taken into account. For the valence
$p\frac{3}{2}$ neutrons in $^{5,6}$He, the WS parameters are the same as those
used above for the $p\frac{3}{2}$ component in the ``core + valence neutron"
wave function of $^{5,6}$He. This choice of the $^6$He g.s. density was made in
a recent study \cite{Kho04} of the interaction cross section induced by $^6$He
beams at high energies. In addition to the IPM density, we have also used the
$^6$He density calculated in a realistic three-body model \cite{Al96} and both
densities give actually the same CRC results.

\begin{figure}[htb]
\hspace*{1cm} \vspace{-1.5cm}
\includegraphics[angle=0,scale=0.6]{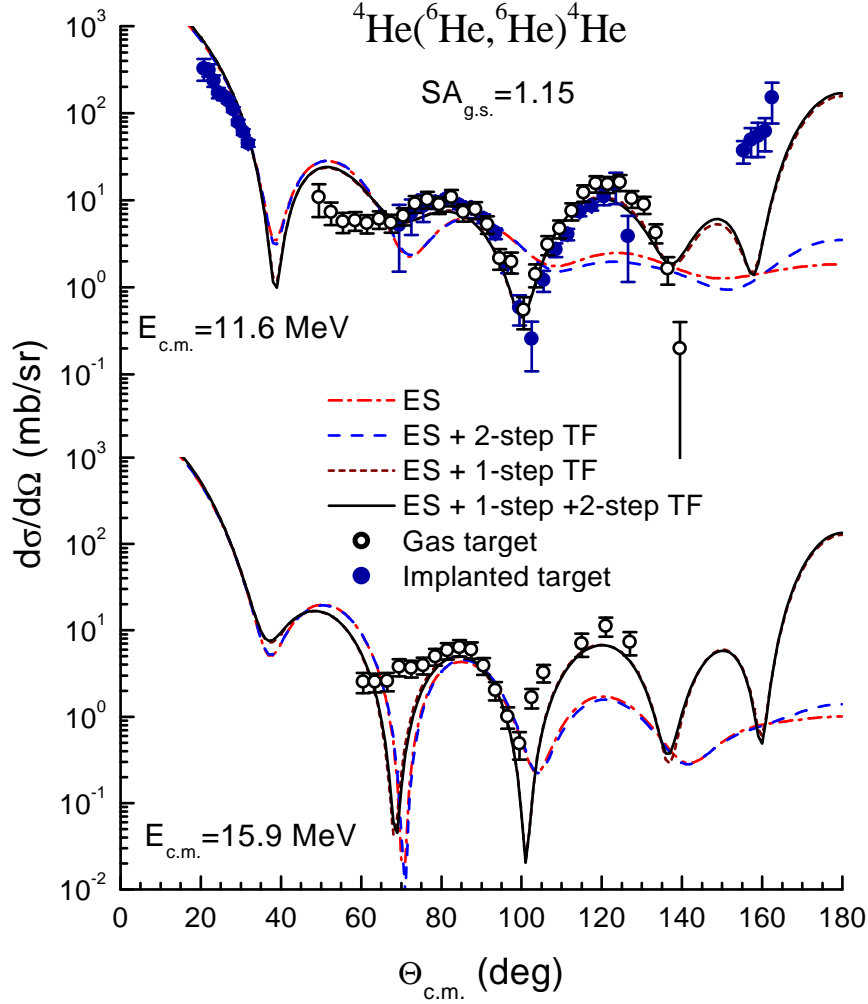}
\caption{Results of CRC calculations, with the direct (one-step) and sequential
(two-step) $2n$-transfer (TF) amplitudes added coherently to that of the elastic
scattering (ES), in comparison with the elastic $^4$He($^6$He,$^6$He)$^4$He data
measured at $E_{\rm c.m.}=11.6$ and 15.9 MeV by Raabe {\sl et al.}
\cite{Raa99,Raa03}. The di-neutron spectroscopic amplitude SA$_{\rm g.s.}=1.15$
was used.} \label{f1}
\end{figure}

To complete the CRC input, one needs to give explicitly the spectroscopic
amplitudes (SA) of one- and two-neutron configurations that enter the overlaps
$<\Psi_{^4{\rm He}}|\Psi_{^5{\rm He}}>$, $<\Psi_{^5{\rm He}}|\Psi_{^6{\rm
He}}>$, $<\Psi_{^4{\rm He}}|\Psi_{^6{\rm He}}>$, and the corresponding transfer
vertices. Without the coupling to the 2$^+$ excitation of $^6$He, the most
sensitive to the $^4$He($^6$He,$^6$He)$^4$He data is the di-neutron
spectroscopic amplitude of $^6$He in the ground state (SA$_{\rm g.s.}$) and it
has been adjusted in each case to reproduce the large-angle data points. Other
SA values do not affect the calculation strongly and were kept unchanged as
taken from Ref.~\cite{Kro01}.

\begin{figure}[htb]
\hspace*{1cm} \vspace{-5cm}
\includegraphics[angle=0,scale=0.6]{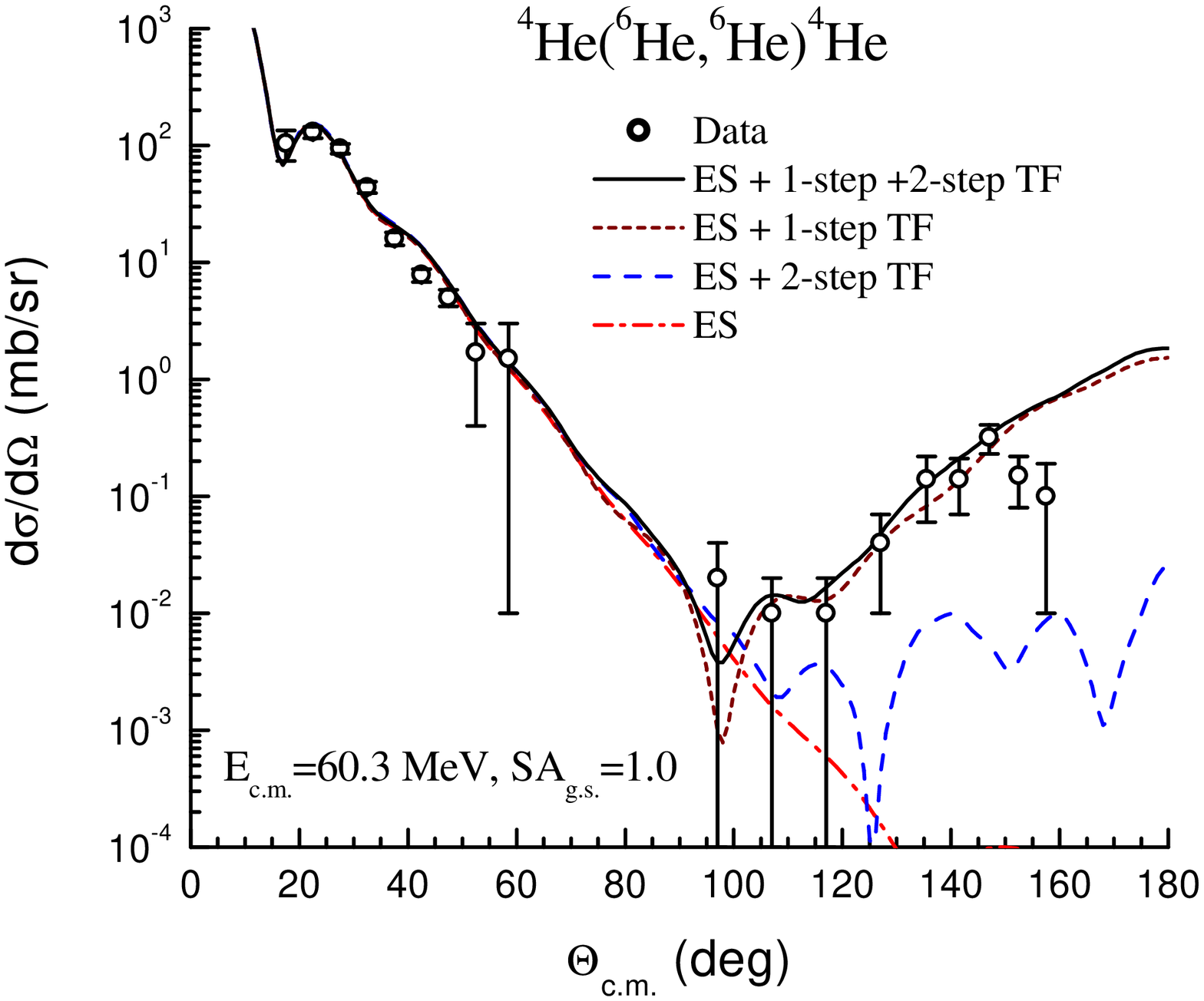}
\caption{The same as Fig.~\ref{f1} for $E_{\rm c.m.}=60.3$ MeV, but with the
di-neutron spectroscopic amplitude SA$_{\rm g.s.}=1.0$, in comparison with the
data obtained by Ter-Akopian {\sl et al.} \cite{Ter98}.} \label{f2}
\end{figure}

Results of our CRC calculations, which take into account two-way coupling
between all the considered channels of the $^4$He($^6$He,$^6$He)$^4$He reaction
at $E_{\rm c.m.}=11.6$ and 15.9 MeV, are plotted in Fig.~\ref{f1}. They show
that the interference between the elastic scattering and 2$n$-transfer
amplitudes gives rise to the observed oscillations of the cross section. One can
also see that the direct one-step transfer is dominant and contribution from the
sequential two-step transfer can be neglected at these low energies. The deepest
minimum of the cross section at $\Theta_{\rm c.m.}\approx 100^\circ$ (best seen
in the data at 11.6 MeV) is due mainly to the direct transfer. The data at 15.9
MeV are of much poorer quality, but they have a minimum at about the same angle,
which is reproduced in our CRC calculation by the same interference mechanism.
In difference from our results, the CDCC calculation of elastic $^6$He+$^4$He
scattering \cite{Rus00} (which reproduces the rise of cross section at backward
angles by taking into account only the coupling between the elastic scattering
and $^6$He breakup channels) does not describe consistently the observed
oscillation pattern (see Fig.~4 in Ref.~ \cite{Rus00}). Moreover, adding the
transfer amplitude to the elastic scattering was reported to deteriorate the
CDCC description of the data. We conclude, therefore, that the $2n$-transfer is
the main physics process responsible for the rise and broad oscillations of the
$^4$He($^6$He,$^6$He)$^4$He cross section at backward angles.

The $^4$He($^6$He,$^6$He)$^4$He data at $E_{\rm c.m.}=60.3$ MeV are compared
with the CRC results in Fig.~\ref{f2}. In difference from the data at lower
energies, these data consist of two parts: the data points at forward angles
which are purely elastic scattering events and those at backward angles which
are due entirely to the 2$n$-transfer process. The OP parameters at $E_{\rm
c.m.}=60.3$ MeV were chosen, therefore, to reproduce data points at forward
angles only. Our scenario for the di-neutron transfer becomes more convincing
after the backward part of the data at 60.3 MeV is well reproduced by the
coupling potential (\ref{e1}) obtained with the same $\Psi_{^6{\rm He}}$
structure as that used at lower energies. Although the contribution from the
sequential transfer becomes more sizable at 60.3 MeV, the direct transfer
remains dominant and, given a rather high experimental uncertainty, one can
still neglect the sequential transfer and assume a direct transfer mechanism at
this energy. Different choices of the WS imaginary OP might give different
cross-section shapes, but the magnitude of the sequential transfer is always
negligible compared to that of the direct transfer. This effect is common for
three considered energies and is likely due to the unbound nature of
$^5$He+$^5$He system, since a negative $Q$-value for breaking the neutron pair
in $^6$He makes the two-step transfer less probable \cite{vOe85}. In a
semiclassical consideration \cite{Sat83}, a negative $Q$-value might also narrow
the ``window" open for the multi-step transfer.

\begin{figure}[htb]
\hspace*{1cm} \vspace{0cm}
\includegraphics[angle=0,scale=0.6]{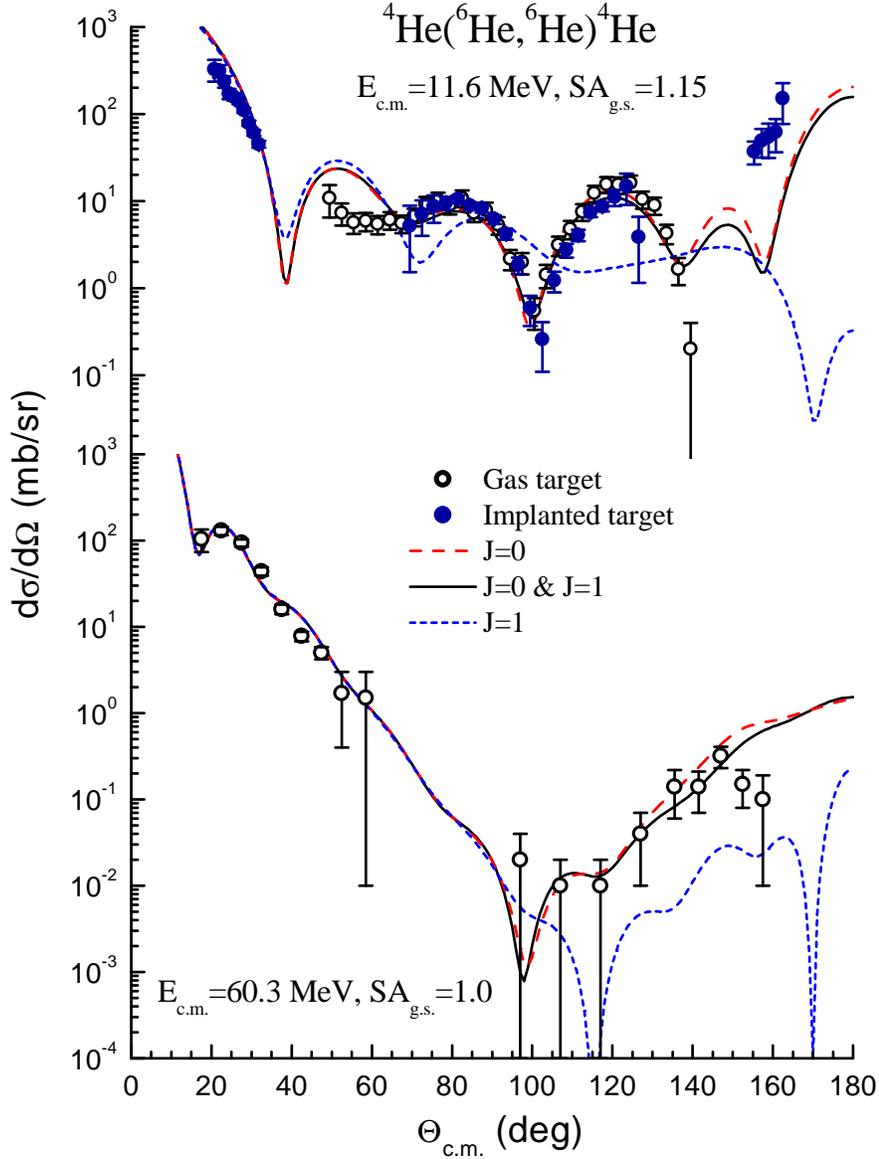}
\caption{Results of CRC calculations, with the direct (one-step) $2n$-transfer
amplitude added coherently to that of the elastic scattering, in comparison with
the $^4$He($^6$He,$^6$He)$^4$He data measured at $E_{\rm c.m.}=11.6$ and 60.3
MeV by Raabe {\sl et al.} \cite{Raa99,Raa03} and Ter-Akopian {\sl et al.}
\cite{Ter98}, respectively. Contributions by the $S$-wave ($L=J=0$) and $P$-wave
($L=J=1$) parts of $^6$He wave function (\ref{wf0}) to the direct transfer are
shown explicitly.} \label{f3}
\end{figure}

%Since the sequential two-step and direct one-step $2n$-transfers are the
%indicators of the ($n-^5$He)- and (2$n-^4$He)-type binding in $^6$He,
%respectively, our result shows that the 2$n-^4$He configuration strongly
%dominates over $n-^5$He configuration in $^6$He.

As already mentioned above, the $P$-wave part of the $^6$He wave function
(\ref{wf0}) belongs to the ``cigar"-type configuration \cite{Tim01}, and it is
necessary to check relative contributions by the $S$- and $P$-wave components of
(\ref{wf0}) to the $2n$-transfer before concluding about the ``di-neutron"
configuration. We have, therefore, made further CRC analysis by omitting either
the $S$- or $P$-wave part of (\ref{wf0}) in the calculation of the coupling
potential (\ref{e1}). Such a decomposition of the direct $2n$-transfer into the
contributions by the $S$- and $P$-waves is shown in Fig.~\ref{f3}. One can see
that the $P$-wave contribution is indeed negligible if it contributes up to 15\%
to the norm of $\Psi_{^6{\rm He}}$.

Given the dominance of the direct $2n$-transfer, coupling potential (\ref{e1})
needs to be evaluated as accurately as possible using realistic $V_{n-\alpha}$
and $^{6,4}$He+$^4$He OP's. We tried first to fit the forward part of the
$^4$He($^6$He,$^6$He)$^4$He data at 60.3 MeV within the OM, using the real
folded potential $V_{\rm Fold}$ and WS imaginary potential. The OM results shown
in Fig.~\ref{f2} were obtained with $V_{\rm Fold}$ renormalized by a factor
$N_{\rm V}\approx 1.15$ and a WS imaginary potential close to that found in
Ref.~\cite{Kro01}. Such a $N_{\rm V}$ factor agrees reasonably well with the
folding analysis of $\alpha$-nucleus elastic scattering \cite{Kho01}, where a
factor of $N_{\rm V}\approx 1.1$ was needed for light targets. We decided,
therefore, to use the same $N_{\rm V}$ factor at lower energies and adjust the
WS parameters of the imaginary OP to fit the $^4$He($^6$He,$^6$He)$^4$He data
over the whole angular range by the CRC results. The volume integral $J_{\rm V}$
of the \emph{renormalized} folded potential turned out to be closer to the
global systematics established for $^4$He+$^4$He \cite{Rao00} than to that for
light heavy-ion systems \cite{Bra97}. For consistency, we have also used $N_{\rm
V}=1.15$ for the folded $^{4,5}$He+$^{4,5}$He potentials. Using a OP systematics
from Ref.~\cite{Rao00}, we found that the contribution from
$^{4,5}$He+$^{4,5}$He imaginary OP's is small and, therefore, was not included
in the final CRC calculation. Concerning the core-core potential, we have used
at the two low energies also a \emph{deep} $^4$He+$^4$He potential \cite{Buc77}
(representing microscopic results of the resonating group method) and a
\emph{shallow} potential parametrized by Ali and Bodmer \cite{Ali66}. These two
potentials were known to give equally good description of $^4$He+$^4$He phase
shifts at low energies, and the CRC results given by them are almost
indistinguishable from those given by the folded $^4$He+$^4$He potential. The
stability of our CRC results with respect to the choice of different core-core
potentials as well as the dominance of the direct $2n$-transfer show
consistently the ``core + 2$n$-cluster" structure of $^6$He.

\begin{figure}[htb]
\hspace*{1cm} \vspace{0cm}
\includegraphics[angle=0,scale=0.6]{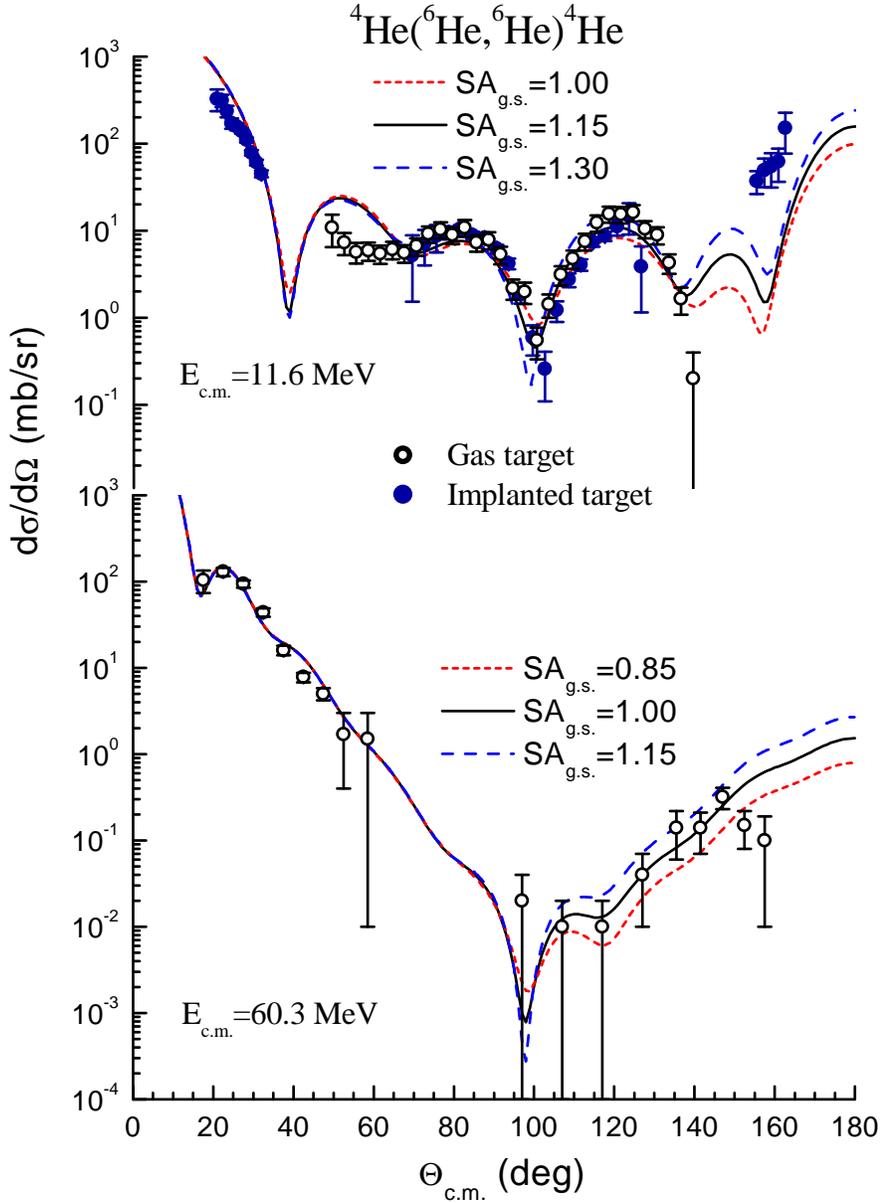}
\caption{Results of CRC calculations, with the direct (one-step) $2n$-transfer
amplitude added coherently to that of the elastic scattering, in comparison with
the $^4$He($^6$He,$^6$He)$^4$He data measured at $E_{\rm c.m.}=11.6$ and 60.3
MeV by Raabe {\sl et al.} \cite{Raa99,Raa03} and Ter-Akopian {\sl et al.}
\cite{Ter98}, respectively. Different di-neutron spectroscopic amplitudes
SA$_{\rm g.s.}$ were used in the calculation of the coupling potential
(\ref{e1}).} \label{f4}
\end{figure}

With a dominance of the direct $2n$-transfer, especially at low energies, a
reliable estimate for the spectroscopic amplitude SA$_{\rm g.s.}$ of the
$2n$-configuration in $^6$He$_{\rm g.s.}$ can be made. In a simple shell model
limit, with the ``core + $2n$-cluster" wave function (\ref{wf0}) normalized to
unity, SA$_{\rm g.s.}$ is expected to be unity if the transfer process exhausts
all the available $2n$-strength in the total wave function. However, SA$_{\rm
g.s.}$ was shown by Timofeyuk \cite{Tim01} to be increased by a factor of
(25/16)$^{1/2}=1.25$ if one takes into account explicitly the center-of-mass
motion as well as the antisymmetrization between all individual nucleons
(including those in $^4$He core). In the present work, we have treated SA$_{\rm
g.s.}$ as a parameter to be found from the best fit to the transfer data by the
CRC results. From the results obtained for the low energies of 11.6 and 15.9 MeV
we can deduce the best fit value SA$_{\rm g.s.}\approx 1.15\pm 0.15$, while the
CRC fit to the data at 60.3 MeV gives SA$_{\rm g.s.}\approx 1.00\pm 0.15$ (see
Table~\ref{t1} and Fig.~\ref{f4}).

\begin{figure}[htb]
\hspace*{1cm} \vspace{0cm}
\includegraphics[angle=0,scale=0.6]{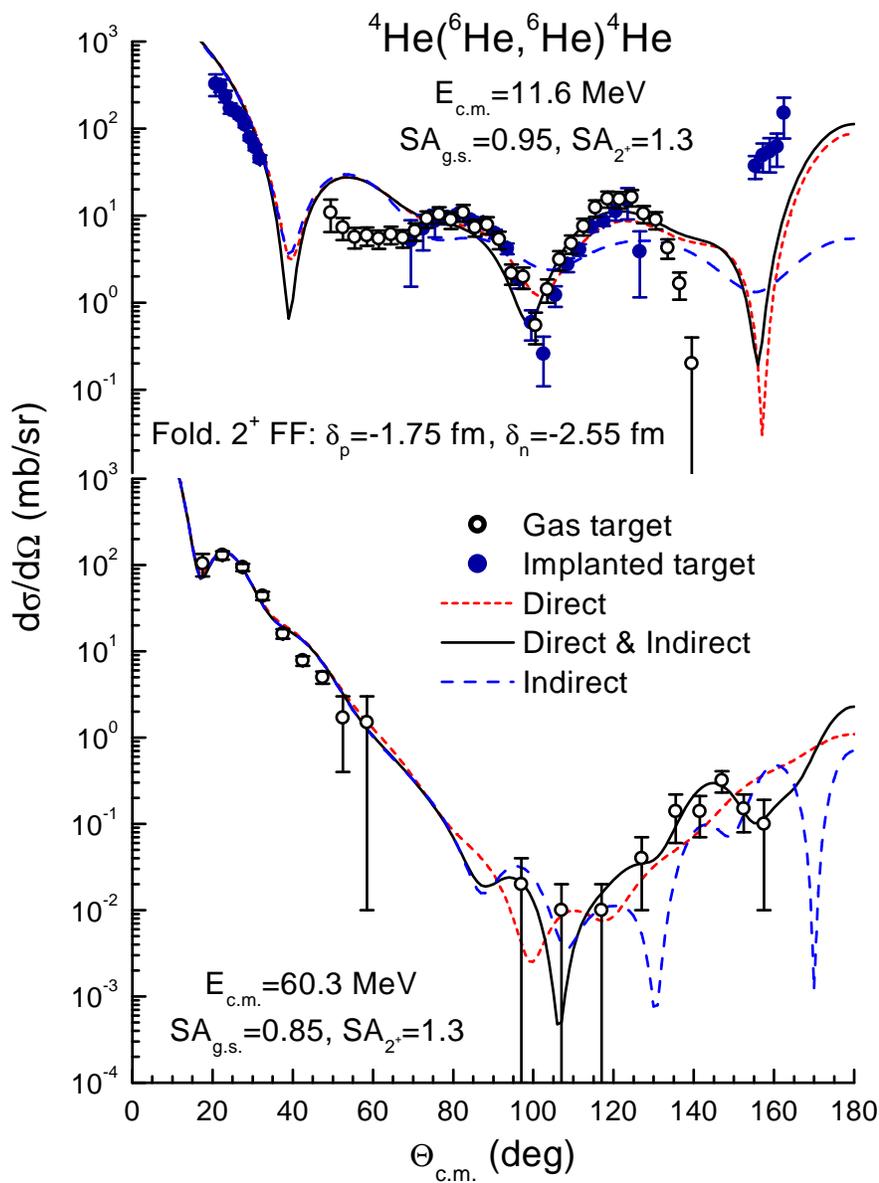}
\caption{Results of CRC calculations, with the $2n$-transfer amplitude added
coherently to that of the elastic scattering, in comparison with the
$^4$He($^6$He,$^6$He)$^4$He data measured at $E_{\rm c.m.}=11.6$ and 60.3 MeV
by Raabe {\sl et al.} \cite{Raa99,Raa03} and Ter-Akopian {\sl et al.}
\cite{Ter98}, respectively. Contributions by the direct and indirect (via 2$^+$
excitation of $^6$He) $2n$-transfers are shown explicitly (see text for more
details).} \label{f5}
\end{figure}

As noted above, the indirect $2n$-transfer route via the 2$^+$ excitation of
$^6$He needs also to be investigated before making conclusion about the
di-neutron configuration in $^6$He$_{\rm g.s.}$. The coupling between the
$^6$He$_{\rm g.s.}+^4$He and $^6$He$^*_{2^+}+^4$He partitions was investigated
earlier in the CDCC analysis of elastic $^6$He$+^4$He scattering \cite{Rus00}
and CRC study of the $^4$He($^6$He,$^6$He)$^4$He reaction  at 60.3 MeV
\cite{Kro01}. While suggesting different scenarios for the
$^4$He($^6$He,$^6$He)$^4$He reaction, these works do show that the (g.s.
$\leftrightarrow 2^+$) coupling is strong and affects significantly the
calculated elastic cross section. Actually, the second-order coupling effect in
this case is two-fold and due to

i) inelastic scattering $^6$He$_{\rm g.s.}\to ^6$He$^*_{2^+} \to
 ^6$He$_{\rm g.s.}$,

ii) indirect $2n$-transfers like $^6$He$_{\rm g.s.}\to ^6$He$^*_{2^+}\to ^4$He
or $^4$He$\to ^6$He$^*_{2^+} \to ^6$He$_{\rm g.s.}$.

As a result, the change in the $^4$He($^6$He,$^6$He)$^4$He cross section is a
mixed effect of the changes caused by both the couplings (i) and (ii), and one
needs to separate the coupling (i) from the total coupling in order to probe the
strength of the indirect $2n$-transfer. Since such an analysis has not been done
sofar, we present here our first attempt to address this interesting problem
using a reasonable choice of the inelastic scattering form factor (FF). We have
first added the inelastic scattering $^4$He($^6$He,$^6$He$^*)^4$He channel and
refitted parameters of the WS imaginary OP (Table~\ref{t1}) to obtained about
the same elastic scattering cross section at forward angles with the CRC
calculation including the elastic, inelastic 2$^+$ scattering and direct
$2n$-transfer only. The DFM \cite{Kho00} was used to calculate the real
inelastic scattering FF, which has been proven to be more accurate than the
standard collective-model FF \cite{Kho00,Be96}. The proton and neutron parts of
the ($^6$He$_{\rm g.s.}\to ^6$He$^*_{2^+}$) transition density used in the DFM
calculation was given by the Bohr-Mottelson prescription \cite{Bo75} of
deforming proton and neutron parts of the $^6$He g.s. density with the
deformation lengths $\delta_{2^+}^{(p)}$ and $\delta_{2^+}^{(n)}$, respectively.
While the proton deformation length can be fixed \cite{Kho00} at
$\delta_{2^+}^{(p)}\approx 1.75$ fm by the empirical transition rate of
$B(E2)\approx 3.2$ e$^2$fm$^4$ \cite{Aum99}, $\delta_{2^+}^{(n)}$ is not
necessarily to be the same. For example, $\delta_{2^+}^{(n)}$ was found to be
more than double $\delta_{2^+}^{(p)}$ for the lowest $2^+$ excitation in
$^{20}$O \cite{Kho03}. To get a realistic estimate of $\delta_{2^+}^{(n)}$ using
the prescription suggested in Ref.~\cite{Kho03}, we have performed a coupled
channel analysis of the inelastic scattering $p$($^6$He,$^6$He$^*_{2^+})p$ data
at 41 MeV/nucleon \cite{Lag01}. We found that $\delta_{2^+}^{(n)}\approx 2.55$
fm which is about 50\% stronger than $\delta_{2^+}^{(p)}$ and, thus, indicates a
strong contribution by the valence neutrons to the 2$^+$ excitation of $^6$He.
The imaginary part of the inelastic FF was given by deforming the WS imaginary
OP with an \emph{isoscalar} deformation length $\delta_{2^+}^{(0)}\approx 2.35$
fm, which was directly obtained \cite{Kho03} from the above values of
$\delta_{2^+}^{(p)}$ and $\delta_{2^+}^{(n)}$. In this way, we have gained
control of some important parameters for our CRC analysis which, otherwise, can
be obtained only when the (purely) inelastic $^6$He+$^4$He scattering data are
available.

After the coupling (i) is properly taken into account, we could estimate the
coupling strength (ii) by adding the indirect $2n$-transfer channel to the CRC
calculation using the new imaginary OP. The same structureless $2n$-cluster in
a (\emph{quasi-bound}) $N=0,\ L=2$ state as that used in Ref.~\cite{Kro01} has
been assumed for $^6$He$^*_{2^+}$. The only remaining parameter is the
di-neutron spectroscopic amplitude SA$_{2^+}$ of $^6$He$^*_{2^+}$, which
determines the strength of the indirect $2n$-transfer. Since the largest-angles
data points at 60.3 MeV were shown in Ref.~\cite{Kro01} to be strongly
sensitive to the indirect $2n$-transfer and could not be described by the CRC
calculation including the coupling (i) only, we have adjusted SA$_{2^+}$ to
reproduce these data points when both the couplings (i) and (ii) were included.
As a result, a perfect agreement with the large-angle data at 60.3 MeV was
obtained with SA$_{2^+}\approx 1.3$ (see Fig.~\ref{f5}). With the direct and
indirect transfer routes now included in equal footing, SA$_{\rm g.s.}$ has
also been readjusted for an optimal agreement of the final CRC results with the
data and it turned out to be slightly reduced (SA$_{\rm g.s.}\approx 0.95$ and
0.85 at 11.6 and 60.3 MeV, respectively). To test the sign effect, we have
performed the CRC analysis using two sets of the inelastic FF's which were
given by the \emph{positive} and \emph{negative} deformation lengths. We found
that the $^4$He($^6$He,$^6$He)$^4$He data at 60.3 MeV is much better reproduced
with the negative deformation ($\delta_{2^+}^{(p)}=-1.75$,
$\delta_{2^+}^{(n)}=-2.55$ and $\delta_{2^+}^{(0)}=-2.35$ fm). Such a
preference of the negative deformation is in an agreement with the earlier CRC
analysis \cite{Kro01} of the same data, but using a simpler ansatz for the
inelastic FF. This sign effect is especially important when the interference
between the direct and indirect transfer amplitudes is not negligible
\cite{Asc84}. In fact, this interference was shown here to be vital for the
full agreement of the CRC results with the data at 60.3 MeV. While the
sequential ($^6$He,$^5$He;$^5$He,$^4$He) transfer is negligible, one can see
from the relative contributions by the direct and indirect transfers shown in
Fig.~\ref{f5} that the indirect ($^6$He,$^6$He$^*_{2^+};^6$He$^*_{2^+},^4$He)
transfer via the $2^+$ excitation of $^6$He is not negligible. It should be
noted that the strength of the indirect transfer (SA$_{2^+}\approx 1.3$) has
been fixed by the CRC fit to the last 3 data points at 60.3 MeV. A future
measurement of the inelastic $^4$He($^6$He,$^6$He$^*_{2^+})^4$He reaction at
about the same energies would be very valuable for a more accurate estimate of
SA$_{2^+}$ from the direct $^6$He$^*_{2^+}\leftrightarrow ^4$He transfer
process. The most realistic values of the di-neutron spectroscopic amplitudes
should be deduced from a CRC analysis including both the direct and indirect
$2n$-transfers if the indirect transfer is significant. From the present CRC
results we can suggest SA$_{\rm g.s.}\approx 0.9\pm 0.1$ and SA$_{2^+}\approx
1.3\pm 0.1$. In this case, instead of the structure model for $^6$He used by
Oganessian {\sl et al.} \cite{Oga99}, one should use a more consistent
structure model for both $^6$He$_{\rm g.s.}$ and $^6$He$^*_{2^+}$ in the CRC
analysis of the $^4$He($^6$He,$^6$He)$^4$He reaction for a direct probe of the
di-neutron and cigar configurations in the $^6$He wave function.

In conclusion, a microscopic CRC analysis of the $^4$He($^6$He,$^6$He)$^4$He
reaction at $E_{\rm c.m.}=11.6,$ 15.9, and 60.3 MeV which includes coherently
the pure elastic scattering, direct (one-step) and sequential (two-step)
2$n$-transfer processes, has been performed for the first time using
semi-microscopic optical potentials for the involved He-He systems and
reasonable choices of the ``core + valence neutrons" wave functions of
$^{5,6}$He. Our analysis showed consistently that the $2n$-transfer is the main
physics process responsible for the rise of the elastic $^6$He+$^4$He cross
section at large angles. The direct $2n$-transfer was found to be dominating
over the sequential transfer and due mainly to the contribution from the
$S$-wave component of the $^6$He wave function. We found further some
indication that the indirect $2n$-transfer via the 2$^+$ excitation of $^6$He
is significant, especially at $E_{\rm c.m.}=60.3$ MeV, and the most reliable
estimate for the di-neutron spectroscopic amplitudes SA$_{\rm g.s.}$ and
SA$_{2^+}$ can be made only if the interference between the direct and indirect
$2n$-transfer amplitudes are taken into account properly. The dominance of the
(direct + indirect) $2n$-cluster transfer shows a strong $2n$-correlation at
the nuclear surface and, thus, confirms the dominance of the 2$n-^4$He
configuration over the $n-^5$He one in the $^6$He wave function.

We thank Gerd Bohlen for helpful discussions and the authors of
Refs.~\cite{Ter98,Raa99,Raa03} for making available the
$^4$He($^6$He,$^6$He)$^4$He data in the tabulated form. D.T.K. is grateful to
the Alexander-von-Humbodlt Stiftung of Germany for the support during his stay
at HMI Berlin in summer 2003, when this work was initiated.

\end{document}